\documentclass[a4paper,noarxiv,twocolumn]{quantumarticle}
\usepackage[utf8]{inputenc}
\usepackage[english]{babel}
\usepackage[T1]{fontenc}

\usepackage{lipsum}
\usepackage{mdframed}
\usepackage{booktabs}
\usepackage{longtable}

\usepackage[numbers,sort&compress]{natbib}

\usepackage{amsmath,amsfonts,bm}

\def\eqref#1{equation~\ref{#1}}

\def\1{\bm{1}}

\DeclareMathAlphabet{\mathsfit}{\encodingdefault}{\sfdefault}{m}{sl}
\SetMathAlphabet{\mathsfit}{bold}{\encodingdefault}{\sfdefault}{bx}{n}

\DeclareMathOperator*{\argmax}{arg\,max}

\usepackage{graphicx}
\usepackage{algorithm}
\usepackage{algpseudocode}
\usepackage{braket}
\usepackage{bbm}

\newtheorem{theorem}{Theorem}[section]
\newtheorem{lemma}{Lemma}[section]
\newtheorem{definition}{Definition}[section]

\newcommand{\PP}{\mathbb{P}}
\newcommand{\RR}{\mathbb{R}}
\newcommand{\EE}{\mathbb{E}}

\usepackage{hyperref}
\usepackage{url}

\title{Provably Robust Training of Quantum Circuit Classifiers Against Parameter Noise}

\author[1]{Lucas Tecot}
\author[2]{Di Luo}
\author[1]{Cho-Jui Hsieh}
\affil[1]{Computer Science, University of California Los Angeles}
\affil[2]{Electrical and Computer Engineering, University of California Los Angeles}

\begin{document}

\maketitle

\begin{abstract}
Advancements in quantum computing have spurred significant interest in harnessing its potential for speedups over classical systems. However, noise remains a major obstacle to achieving reliable quantum algorithms. In this work, we present a provably noise-resilient training theory and algorithm to enhance the robustness of parameterized quantum circuit classifiers. Our method, with a natural connection to Evolutionary Strategies, guarantees resilience to parameter noise with minimal adjustments to commonly used optimization algorithms. Our approach is function-agnostic and adaptable to various quantum circuits, successfully demonstrated in quantum phase classification tasks. By developing provably guaranteed optimization theory with quantum circuits, our work opens new avenues for practical, robust applications of near-term quantum computers.
\end{abstract}

\section{Introduction}

In the past few decades, the field of quantum computing has grown dramatically \citep{farhi_quantum_2014, harrigan_quantum_2021, moll_quantum_2017, klco_quantum-classical_2018, peruzzo_variational_2014}. Enticed by the potential of speed-ups over classical computers, great efforts have been devoted to not only building quantum computers \citep{bluvstein_logical_2024}, but also exploring how to achieve these speed-ups when the proper devices exists \citep{biamonte_quantum_2017} with a wide variety of approaches, from carefully-crafted algorithms \citep{dalzell_quantum_2023} to data-learned optimized approaches \citep{cerezo_variational_2021}. 

However, all existing quantum algorithms suffer from the existence of noise on the current quantum hardware. Until the theory and practice of error-corrected quantum computing makes significant progress, all quantum algorithms need to find ways to be robust to the noise that exists in noisy intermediate-scale quantum (NISQ) computers \citep{preskill_quantum_2018}. Furthermore, there are many types of noise, from noise inherent to quantum systems that can't be fully isolated from their environment, to simple instrumentation noise that occurs whenever preparing a continuously-parameterized operation \citep{saki_impact_2021, cai_quantum_2023}. As a result, a number of works have explored different ways to fight noise in the NISQ era \citep{west_towards_2023, weber_optimal_2021, weber_toward_2022}.

In parallel, significant effort has been made to improve the noise robustness of machine learning models. One strong method to achieve this is randomized smoothing, which certifies a "smoothed" classifier’s robustness to input perturbations by analyzing the distribution of outputs under noise \citep{cohen_certified_2019, yang_randomized_2020, levine_robustness_2019, tecot_robustness_2021, salman_provably_2020, salman_denoised_2020}. 

In this work, we connect tools from classical machine learning to the quantum setting and offer another tool to fight against noise for algorithms deployed on NISQ devices. Specifically we tackle parameter gate noise, in which the prepared parameters of a continuously-parameterized quantum gate may be different than it is intended to be. Our method is a \textit{certified robustness} method, in which we can guarantee a certain level of noise will not affect the end result of the quantum process. We highlight our contributions as follows:
\begin{enumerate}
    \item Our work provides a provably guaranteed framework and theory for training parameterized quantum circuit classifier under parameter noise. While most other methods consider adversarial attacks on inputs or mid-circuit noise inherent to quantum systems, we explicitly tackle the use case that accounts for instrumentation error in devices. As long as the desired noise-model and algorithm can be expressed as a parameterized circuit, our method can be used to find an optimal robustness certificate for changes in those parameters. 
    \item Our method is simple, easy to deploy, and naturally connected to Evolutionary Strategies (ES), an optimization algorithm already commonly used by the variational quantum algorithm (VQA) community \citep{gil-fuster_understanding_2023, anand_natural_2021}. Due to its ability to avoid computing costly quantum gradients \citep{abbas_quantum_2023}, a practitioner can achieve robustness certificates using our method with minimal effort. Furthermore, our method can be applied on top of any error-mitigation method, allowing to be easily combined with other methods to further boost the robustness they provide.
    \item Our approach is successfully demonstrated on quantum phase classification tasks. Our results present clear robustness-variance trade-off and a robustness-variance correlation, which provide insights and understanding on the sensitivity of the parameters in quantum parameterized circuits.
\end{enumerate}

In summary, we develop a provably noise-resilient training theory and algorithm for parameterized quantum circuits. With its flexibility and ease of adaptation, our approach achieves noise robustness, paving the way for practical applications on near-term quantum computers.

\section{Method} \label{sec:method}

\subsection{Noise-resilient Theorem}

\begin{definition} [PQC Classifier]
A classifier $C$ is called a parameterized quantum circuit (PQC) classifier if it is constructed in the following way:
\begin{gather}
    C(\theta, x)_i = \bra{\phi_0} V^\dag(x) U^\dag(\theta) A_i U(\theta) V(x) \ket{\phi_0} \\
    U(\theta) = U_L(\theta_L) \cdots U_2(\theta_2) U_1(\theta_1) \\
    U_l(\theta_l) = \prod_{m} e^{-i \theta_m H_m} W_m
\end{gather}
    
\end{definition}
where $C(\theta, x)_i$ is the quantum classifier's probability assigned to class $i$, $A_i$ are easily measurable observables, $V(x)$ is a data-dependent unitary, $W_m$ is an unparameterized unitary, $H_m$ is a Hermitian operator, and $\theta_l$ is the $l$-th element of $\theta$. PQC classifiers have been considered in various quantum machine learning setups \citep{cerezo_variational_2021, schuld_circuit-centric_2020, weber_optimal_2021}. While we follow the above particular form of PQC in this work, our discussion is general as long as the classifier comes from a quantum circuit with unitaries depending on $x$ and $\theta$.

In this work, we focus on noise effect on the given PQC classifier parameters. If the PQC classifier parameters are fully robust to noise, any possible noisy perturbation on the parameters should not change the correct classification of the task. More precisely, we formalize this notion as follows.

\begin{definition}[Noise-resilient PQC Classifier]
A PQC classifier is parameter noise-resilient if the following is true
\begin{gather}
    \forall (x,y) \in D, \\
    \left (\argmax_i [C(\theta, x)_i] = y \right ) \implies \\
    \qquad \left ( \argmax_i [ C(\theta + \delta, x)_i ] = y \right ) 
\end{gather}
    
\end{definition}
for all perturbations $\delta$ sampled from a given domain. 

Our goal is to develop robust training theory and algorithms for PQC classifier so that it is provably parameter noise-resilient. To tackle this problem, we integrate approach from classical machine learning with parameterized quantum circuits. More specifically, we develop randomized smoothing certified robustness theory \citep{tecot_robustness_2021} under the setting of PQC classifier. To start with, we introduce the concept of smoothed PQC classifier.

\begin{definition} [Smoothed PQC Classifier]
    Let $C$ be a PQC classifier with possible prediction classes $\gamma=\{1, \dots, N\}$. A smoothed PQC classifier $G_{\sigma}$ is defined as:
    \begin{align}
        G_{\sigma}(\theta, x) = \argmax_{z \in \gamma} \PP \left ( \argmax_{i \in \gamma} \left [ C(\theta+\epsilon, x)_i \right ] = z \right ), 
    \end{align}
    where $\epsilon \sim \mathcal{N}(0, \Sigma)$ and $\Sigma$ is a diagonal matrix with vector $\sigma^{2}$ as the diagonal.
\end{definition}

\begin{theorem} [Noise-resilient Condition] \label{indep_vars_thm}
    Let $C$ be a PQC classifier and $G_{\sigma}$ to be the corresponding smoothed PQC classifier. If $G_{\sigma}(\theta, x) = c_a$, then $G_{\sigma}(\theta+\delta, x) = c_a$ for any $\delta$ vectors that satisfy
    \begin{align}
        \| \delta \oslash \sigma \|_2 &< \frac{1}{2} \big (\Phi^{-1}(p_A) - \Phi^{-1}(p_B) \big )
    \end{align} 
     where $\oslash$ is the Hadamard (element-wise) division, $\|\cdot\|_2$ is the $L_2$ norm, $\Phi^{-1}$ is the inverse of the standard Gaussian CDF, and 
    \begin{align}
        p_A &= \PP \left ( \argmax_{i \in \gamma} [C(\theta+\epsilon, x)_i] = c_a \right ) \\
        p_B &= \max_{c\neq c_a} \PP \left ( \argmax_{i \in \gamma} [C(\theta+\epsilon, x)_i] = c \right )
    \end{align} 
    with $\epsilon \sim \mathcal{N}(0, \Sigma)$ and $\Sigma$ is a diagonal matrix with vector $\sigma^{2}$ as the diagonal.
\end{theorem}

Using the above theorem, by smoothing a given PQC (i.e. re-evaluating the model multiple times, with $\epsilon$ sampled from a multivariate Gaussian each time), we can guarantee with high probability that any noise $\delta$ caused by the environment will not change the prediction results, as long as it satisfies the given bound. All we need to do is estimate bounds on $p_A$ and $p_B$ using concentration inequalities and then we can directly apply the theorem (See Section \ref{sec:prob_est} for more details). Since standard PQC classifiers require multiple evaluations due to the probabilistic nature of quantum measurements, our smoothed PQC classifier operates in practice similarly to a standard PQC classifier in both method and computational cost. The implementation only involves determining a $\sigma$ vector in addition to the ideal $\theta$ parameters for sampling, making it resource-efficient for near-term quantum computers. We further note that the ability of our approach to vary $\sigma$ across different parameters, as opposed to using a uniform robust radius, adds flexibility to enhance the system's resilience to noise.

\subsection{Robust Training Algorithm} \label{sec:optimization}

Theorem.~\ref{indep_vars_thm} has provided a provably guarantee on noise resilience for PQC classifier. Next, we consider how to train our circuits and improve this robust-bound. First, let us consider a commonly used method for training PQCs - Evolution Strategies (ES). This optimization algorithm is commonly used in the quantum community \citep{gil-fuster_understanding_2023, anand_natural_2021} due to its ability to optimize while avoiding computing costly quantum gradients \citep{abbas_quantum_2023}. What ES typically optimizes for is
\begin{gather}
    J = \EE_{(x,y) \in D, \epsilon \sim \mathcal{N}(0, \Sigma)} O(\theta + \epsilon, x, y)
\end{gather}
where $D$ is our training dataset, and $O$ is the objective function.
While the "search distribution" that we sample $\epsilon$ from can vary depending on the version of ES, we consider a multivariate Gaussian which is the distribution commonly used by the most popular versions of ES in PQC optimization, such as CMA-ES and NES \citep{hansen_cma_2006, wierstra_natural_2011}.

\textbf{Connection of ES and Noise-resilient Condition.} We highlight that the equation of ES is closely related to the right-side of the bound in Theorem \ref{indep_vars_thm}. In ES, we optimize the parameters of a multivariate Gaussian to minimize an objective in expectation. For Theorem \ref{indep_vars_thm}, we desire to optimize $\theta$ and $\sigma$ to maximize our robust bound. Note that $\theta + \epsilon$ is also a multivariate Gaussian, where $\theta$ is the mean and $\sigma$ are the independent variances. Therefore, if we simply change the objective function $O$ to calculate the margin of prediction instead, we can exactly optimize for the right-hand side of Theorem \ref{indep_vars_thm} using ES. More precisely, we can re-formulate the optimization goal as: 
\begin{gather}
    \argmax_{\theta, \sigma} \left [ \EE_{(x,y) \in D, \epsilon \sim \mathcal{N}(0, \Sigma)} O(\theta + \epsilon, x, y) \right ] \\
    O(\theta + \epsilon, x, y) = \frac{1}{2} \big ( \Phi^{-1}(p_A) - \Phi^{-1}(p_B) \big )
\end{gather}
where $\Sigma, p_A, p_B$ are all as defined in Theorem \ref{indep_vars_thm}. Since our theorem relies on having independent variances rather than a full covariance matrix, we use sNES \citep{wierstra_natural_2011}, which is a variation on NES that assumes independent variances between input elements.

Additionally, note that while this procedure fits perfectly for optimizing a variational quantum algorithm, it also works for PQC's that have parameters that aren't optimized for. All one needs to do is simply removing the $\theta$ update step for all fixed elements, so the parameter in question remains fixed and we only need to find an ideal $\sigma$ component for that element. Furthermore, we note that our method can be applied to any parameterized quantum system, including those that utilize other error-mitigation methods. This allows us to easily boost robustness by combining with existing methods.
\begin{table}[t]
    \centering
    \begin{tabular}{c}
        \textbf{Training Procedure}  \\
        \hline \rule{0pt}{3ex} 
        \parbox{0.4\textwidth}{
        \begin{algorithmic}
        \For {$N$ iterations}
            \For {$k=1 \dots \lambda$}
                \State $s_k \sim \mathcal{N}(0,I)$
                \State $z_k \gets \theta + \sigma s_k$
                \State $f_k \gets (\Phi^{-1}(p_A) - \Phi^{-1}(p_B)) / 2$
            \EndFor
            \State $s'_k \gets$ Sort all $s_k$ w.r.t. $f_k$ 
            \State $u_k \gets \frac{\max(0, \log(\lambda / 2 + 1) - \log(k))}{\sum_{j=1}^\lambda \max(0, \log(\lambda / 2 + 1) - \log(j))} - \frac{1}{\lambda}$
            \State $\nabla_{\theta} J \gets \sum_{k=1}^\lambda u_k s'_k$
            \State $\nabla_{\sigma} J \gets \sum_{k=1}^\lambda u_k({s'}_k^2 - 1)$
            \State $\theta \gets \theta + \eta_{\theta} \cdot \sigma \cdot \nabla_{\theta} J$
            \State $\sigma \gets \sigma \cdot \exp(\eta_{\sigma} / 2 \cdot \nabla_{\sigma} J)$
            \State $\sigma \gets \sigma + \eta_{r} \cdot \text{reg}(\sigma)$
        \EndFor
        \end{algorithmic} 
        } \\ \hline \rule{0pt}{3ex} \\
         \textbf{Deployed Model Use} \\
         \hline \rule{0pt}{3ex}
        \parbox{0.4\textwidth}{
        \begin{algorithmic}
        \State \textbf{Input :} $x, \theta$
        \For {$k=1 \dots M$}
            \State $s_k \sim \mathcal{N}(0,I)$
            \State $z_k \gets \theta + \sigma s_k$
            \State $c_{k,i} \gets C(z_k, x)_i$
        \EndFor
        \State $p_i \gets \frac{1}{M} \sum_{k=1}^M c_{k,i}$
        \State \textbf{Output :} $\argmax_i p_i$
        \end{algorithmic}
        } \\ \hline \rule{0pt}{3ex}
    \end{tabular}
    \caption{Training and use of our method outlined in Section \ref{sec:method}. All variables are as defined by Theorem \ref{indep_vars_thm}. See Section \ref{sec:regularization} for the possible definitions of the $reg(\cdot)$ function. }
    \label{alg:sNES}
\end{table}

\subsection{Variance Regularization} \label{sec:regularization}

After we have identified how to optimize to improve the right side of the bound in Theorem \ref{indep_vars_thm}, we now turn to improve the left side in the bound. Notice that there is an implicit trade-off occurring in this bound; while the left side always benefits from a larger $\sigma$, making it larger introduces the risk of decreasing the accuracy of the smoothed classifier and in turn decreasing the right side of the bound. As such, to address this trade-off we add regularization into our optimization procedure. This allows us to define an optimization trade-off between improving the left-side of the bound via a large $\sigma$ and the right-hand side by encouraging high accuracy. We control this trade-off with a hyperparameter, and in practice we often sweep over many values of this coefficient to understand the nature of accuracy-robustness trade-off per experiment. To control the magnitude of $\sigma$, we utilize two types of regularization methods, both of which have similar performances in our experiments. (See Section \ref{sec:reg} for more details.)

\section{Metrics} \label{metrics}

To test how robust an approach is under noise, we need to consider proper metric for quantification.
While there are many ways to achieve this, understanding the bound from Theorem \ref{indep_vars_thm} from a geometric perspective is beneficial. 
Note that we can re-arrange the terms of the bound in Theorem \ref{indep_vars_thm} to form the equation of a hyper-
ellipsoid:
\begin{gather} \label{eq:cert_shape}
    \sum_{i=1}^D \frac{\delta_i^2}{(s_e \sigma_i)^2} < 1 \\
    s_e = \frac{1}{2} \left (\Phi^{-1}(p_A) - \Phi^{-1}(p_B) \right ).
\end{gather}

In other words, any perturbation that exists within this hyper-ellipsoid will satisfy the bound and not cause a change in the model's prediction result.

Using this re-formulation, we can use the volume of this hyper-ellipsoid as a metric, as it is generally desirable to be robust to the largest possible space of perturbations. The volume of this hyper-ellipsoid is:
\begin{equation} \label{eq:cert_volume}
    V = \frac{2\pi^{D/2}}{D\Gamma(D/2)}\prod_{i=1}^D s_e \sigma_i.
\end{equation}

Using this geometric understanding, we can finally define a handful of useful metrics that capture a wide range of what most use-cases would desire to optimize for:

\textbf{Certified Area Geometric Mean} : The certified area (Equation \ref{eq:cert_volume}) taken to the power of $\frac{1}{D}$, which is $V^{1/D}$. Since the certified area can be very small and vary wildly depending on the dimensionality of the problem, we use the geometric mean to make it easier to think about and compare from experiment to experiment. Conceptually this can be thought of as if we take the volume of the $D$-dimensional hyper-ellipsoid, re-shape it into a $D$-dimensional cube, and then calculate the length of the sides of the cube.

\textbf{Semi-Axis Average} : The average of the semi-axes of the certified hyper-ellipsoid (Equation \ref{eq:cert_shape}), which is $\overline{s_e \sigma} = \frac{1}{D}\sum_{i=1}^D s_e \sigma_i$. Since conceptually each semi-axis length can be thought of as the maximum one can perturb a given parameter, this metric can be interpreted as the maximum that a parameter can change on average under noise.

\textbf{Semi-Axis Standard Deviation} : The standard deviation of the semi-axes of the certified hyper-ellipsoid, which is $\sqrt{\frac{1}{D} \sum_{i=1}^D (s_e \sigma_i - \overline{s_e \sigma})^2}$. Tracking this metric may provide insights into the sensitivity of the parameters in PQC. (See Section \ref{sec:semi-axis-var}.)

\textbf{Smoothed Accuracy} : The accuracy of the smoothed classifier ($G_{\sigma}$ in Theorem \ref{indep_vars_thm}), which is $\EE_{(x,y) \sim D}  \left [ \mathbbm{1} (G_{\sigma}(\theta, x) = y ) \right ]$. $G_{\sigma}(\theta, x)$ is calculated by sampling the underlying PQC with many different parameter-samples (using the Gaussian found by the process outlined in Section \ref{sec:optimization}) and averaging the resulting probabilities.

In our experiments, all numeric are calculated using an average over a test dataset, none of which are seen during training. Before describing the experiments, we introduce the type of plots we will produce with the results.

\subsection{Robustness-Accuracy Trade-off} \label{sec:robust_acc_front}

As mentioned in Section \ref{sec:regularization}, there often exists a trade-off between accuracy and robustness when training smoothed classifiers. As such, we desire to understand the best trade-off we can achieve. In order to do this, we run a randomized hyperparameter sweep over the method described in Section \ref{sec:method}. Specifically, we modulate over the regularization term strength term (Section \ref{sec:regularization}) and all the hyperparameters of sNES. Looking at each level of smoothed accuracy, we select only the runs that achieve the best robustness metric for said accuracy and plot these points. 

Note that this robustness-accuracy trade-off usually only exists for a small section of runs in the higher-accuracy regime, as any regularization strength coefficient that is too high will cause the accuracy margin of the model to dramatically decrease, which will in turn decrease the robustness metric. To properly understand the relevant frontier of this trade-off, we only plot the points on this frontier, and fit a line to them in order to understand the numerics a practitioner can expect to achieve on a similar problem without needing to do significant tuning. These plots can be seen in the first row of Figure \ref{fig:qb12}.

\subsection{Robustness-Variance Correlation} \label{sec:semi-axis-var}

We also produce plots to illustrate the relationship of robustness metrics (i.e. certified area geometric mean and semi-axis average) versus the semi-axis standard deviation. A high standard deviation indicates that the "robust space" hyper-ellipsoid (Equation \ref{eq:cert_shape}) has some dimensions much longer than others, which indicates some parameters are more susceptible to noise than others. Conversely, a low standard deviation indicates that it is closer to a sphere, which indicates all parameters should tolerate near-equal amounts of noise.

For this analysis, we only include runs that achieve high accuracy, as these are the only points that are relevant to the "robustness-accuracy frontier" described in Section \ref{sec:robust_acc_front}. We plot all runs that achieve above the minimum accuracy shown in the "robustness-accuracy frontier" plots. We then bin these runs to show the mean and standard deviation of the semi-axis standard deviation for all runs that achieve a similar robust metric. Similar to the robustness-accuracy trade-off plots, we also fit a line to these points to understand the overall corelation. These plots can be seen in the last row of Figure \ref{fig:qb12}.

\section{Experiments} \label{sec:experiments}

In this work, we consider phase classification, which can be viewed as a classification task from the machine learning perspective. We chose this task because it is important in condensed-matter physics, and as a result are often used in benchmarks \citep{carrasquilla_machine_2017, broecker_machine_2017}. In these problems, given the ground state quantum state, the objective is to predict what phase of the ground state originates from. For our experiment, we generate mutually exclusive train and test datasets of 50 samples each. The train dataset is used to train the model, and all statistics reported are averages over the entire test set. We do randomized hyperparameter sweeps in order to understand what we can optimally achieve, but in practice most hyperparameters (aside from the regularization strength term) worked well as long as they were within a reasonable range (See Section \ref{sec:hyperparams} for more details).

\subsection{Classification Model}

For our experiments, we use the Quantum Convolutional Neural Network (QCNN) \citep{cong_quantum_2019}. Specifically we use a form of QCNN that uses rotational and control X, Y, and Z gates to implement generic 1 and 2 qubit gates \citep{vatan_optimal_2004, noauthor_quantum_nodate}. We certify for all phase shift noise in any gate that uses a parameter-defined angle.

\subsection{Cluster Phase Classification} \label{sec:phase_class}

We consider the generalized cluster Hamiltonian, which is commonly studied in other works \citep{gil-fuster_understanding_2023, caro_generalization_2022}. The Hamiltonian in this setup is:
\begin{gather}
    H = \sum_{j=1}^n (Z_j + j_1 X_j X_{j+1} - j_2 X_{j-1} Z_j X_{j+1}).
\end{gather}

This Hamiltonian contains 4 phases, depending on the values of the coefficients $j_1$ and $j_2$. The values that belong to each phase are illustrated in Figure 2 of \citet{gil-fuster_understanding_2023} (included in our appendix as Figure \ref{fig:cluster_diagram} for convenience). For these experiments, the ground states are found exactly via diagonalization and loaded directly as the starting state of the circuit. Our data is uniformly sampled from $j_1 \in [-4, 4], j_2 \in [-4, 4]$. We present the results in Figure \ref{fig:qb12} with a Hamiltonian of 12 qubits.

\begin{figure}[t]
\begin{center}
\includegraphics[width=0.5\textwidth]{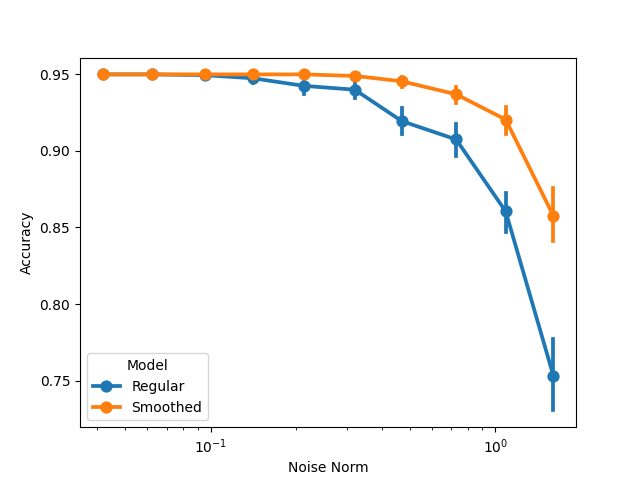}
\end{center}
\caption{Accuracy over 20 test data points for a well-trained PQC and smoothed PQC with varying levels of noise added to the parameters. Each point and bar pair indicates the mean and confidence interval of 100 noisy-parameter samples from a gaussian with variances corresponding to $\sigma$ of the smoothed classifier multiplied by various scaling constants. "Noise Norm" is the average L2-norm of the noise sampled that produced each point.}
\label{fig:compare_plot}
\end{figure}

\begin{figure*}[t]
\begin{center}
\includegraphics[width=0.47\textwidth]{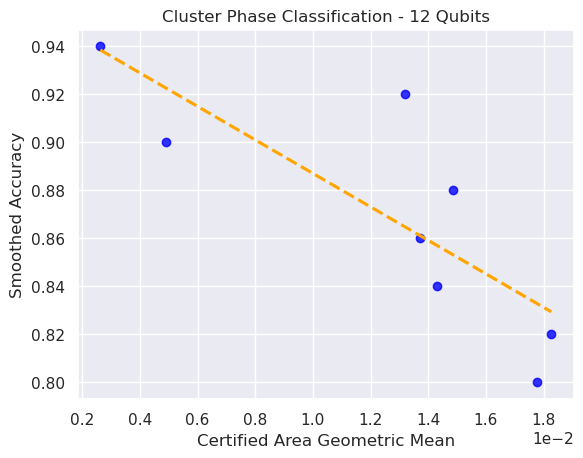}
\includegraphics[width=0.47\textwidth]{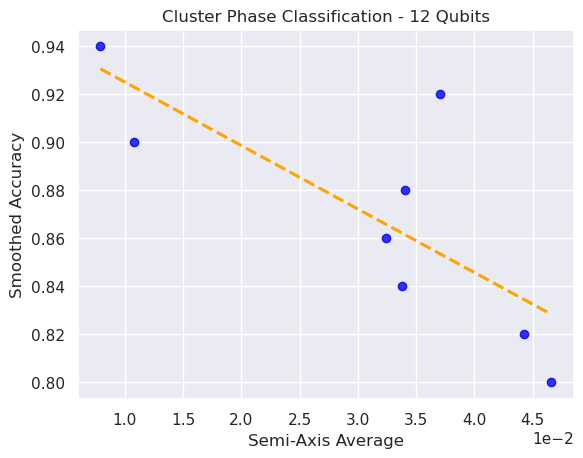}
\includegraphics[width=0.47\textwidth]{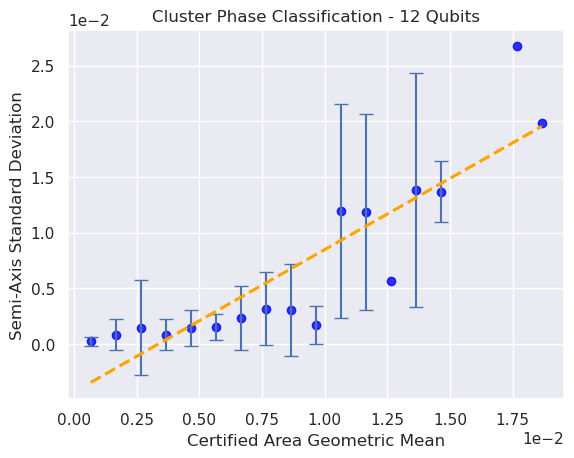}
\includegraphics[width=0.47\textwidth]{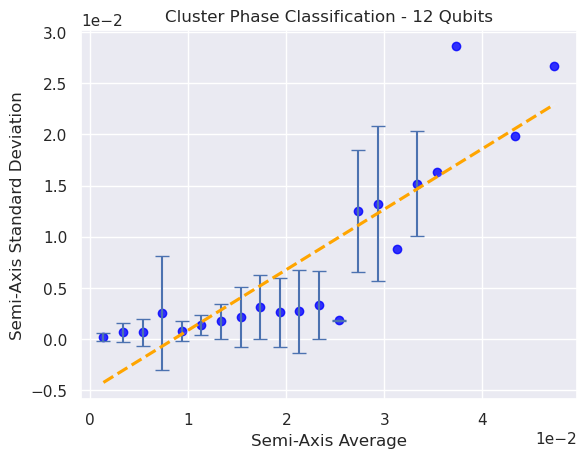}
\end{center}
\caption{Phase classification for the generalized cluster Hamiltonian of 12 qubits, as outlined in Section \ref{sec:phase_class}. The first row illustrates the trade-off between accuracy and robustness, as described in Section \ref{sec:robust_acc_front}. The last row shows the robustness-variance correlation, as described in Section \ref{sec:semi-axis-var}. While our results may vary due to randomness and instability in optimization, we include a linear fit line to indicate the general trend. 
}
\label{fig:qb12}
\end{figure*}

\textbf{Results Discussion :} 
To illustrate the benefit of our method, we first test the robustness of a trained model on real sampled noise. First we take a trained smoothed PQC from our hyperparameter sweep that is able to achieve both accuracy and robustness results that are near the maximum we observed ($88\%$ accuracy on the training test set and certified area geometric mean of $1.309 * 10^{-2}$). We then compare it to a regular PQC that also achieves optimal accuracy on the test set ($92\%$ accuracy). The parameters of each model are then exposed to varying levels of noise. We plot the average accuracy over a test dataset per noise level. These results are shown in Figure \ref{fig:compare_plot}. We demonstrate that a properly trained randomized smoothed model is near-identical to a well-trained regular model for low levels of noise, but achieves better and more robust performance as the noise level increases.

Next we analyze the certified robustness levels we are able to achieve in training. The first row of Figure \ref{fig:qb12} demonstrates the robustness-variance trade-off, where smoothed accuracy decreases with higher robustness metrics. Note that in these experiments, we are able to achieve a certified area geometric mean ranging from roughly $0.002$ to $0.018$, and a semi-axis average ranging from $0.005$ to $0.045$ depending on the smoothed accuracy level. This means that one could likely expect to certify the robustness of similar experiments that experience this amount of phase-shift noise per parameter. While the usefulness of this level of robustness depends on each individual system, noise level, and desired accuracy, it is shown that such robustness could be sufficient for certain near-term systems \citep{bluvstein_quantum_2022, yi_robust_2024, wood_special_2020}. 

Furthermore, there is a clear correlation between high semi-axis standard deviation and our robustness metrics. 
This indicates that different parameters of the PQC have varying amounts of robustness to noise, and as a result we are able to leverage these differences to improve our overall robustness to noise. It illustrates the advantage of our approach with varying $\sigma$ for different parameters compared to a uniform robust radius, which provides more flexibility to be noise-resilient.

\section{Conclusion}

In this work we have developed a provably noise-resilient approach for training parameterized quantum circuit classifiers. Our method is flexible for any quantum circuit and easy to deploy on NISQ quantum device with a natural connection to a Evolutionary Strategies. This makes it extremely simple for practitioners to use our method on any of their existing experiments, both to enhance robustness and to gain insights into the sensitivity of the quantum devices. Future work could explore using the shape of the $\sigma$ vector to understand the sensitivity and importance of parameters in PQC classifiers, especially for different quantum ansatz. Additionally, further optimization and regularization of the PQC classifier could be customized based on specific performance metrics. Expanding the method to include a full-covariance matrix in randomized smoothing could also provide more flexibility in smoothing techniques. Furthermore, it is an open direction to generalize other types of quantum circuits training, such as VQE or QAOA \citep{farhi_quantum_2014}. Our work integrates the frontier machine learning algorithm with quantum computation, opening up opportunities for robust applications of near-term quantum computers.

\clearpage

\bibliography{references}
\bibliographystyle{iclr2025_conference}

\appendix

\clearpage

\onecolumngrid
\begin{center}
    \noindent{\Large\textbf{Supplementary Material}}
	\bigskip
		
	\noindent\textbf{\large{}}
\end{center}

\onecolumn

\section{Additional Experimental Details}

In our cluster phase classification task, there are four phases as the following figure shows
\begin{figure}[H]
\begin{center}
\includegraphics[width=0.5\textwidth]{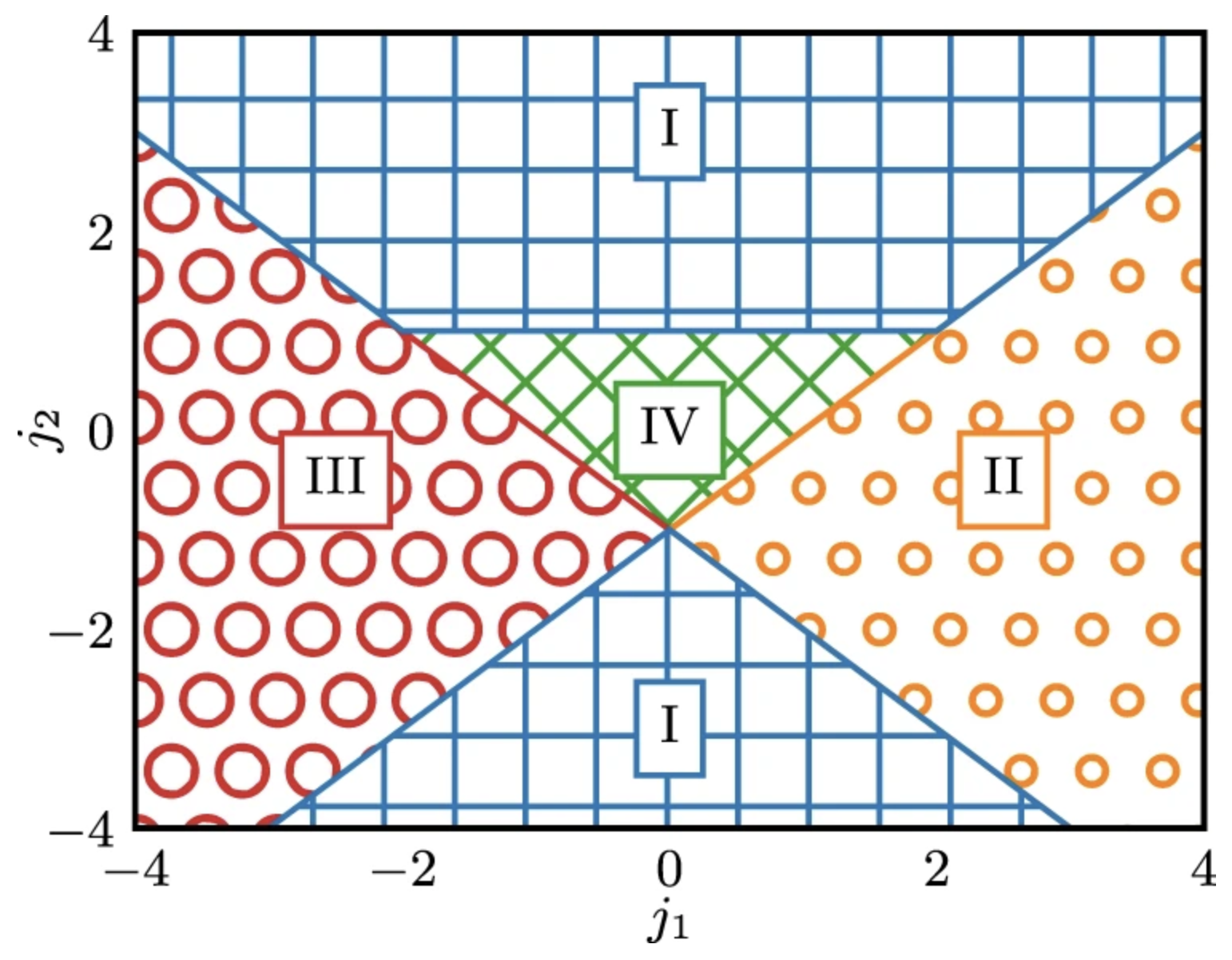}
\end{center}
\caption{Figure 2 from \citet{gil-fuster_understanding_2023}. Illustrates the different phases of the generalized cluster phase-classification problem outlined in Section \ref{sec:phase_class}.}
\label{fig:cluster_diagram}
\end{figure}

\section{Implementation Details}

\subsection{Hyperparameters} \label{sec:hyperparams}

For all experiments we do randomized hyperparameter sweeps in order to understand what we can optimally achieve. Because we are looking at a combination of robustness and accuracy, it is difficult to definitively say which hyperparameters are optimal because it can vary from different levels of accuracy and robustness. That being said, in our sweeps we found that the runs that produced optimal results (near-best robustness metrics for a given accuracy level) came from a wide range of hyperparameter values, and often a hyperparameter would only destroy performance if it was far too low or high. The acceptable range we found for each hyperparameter is as follows. $\sigma_{0}$ is the initial value of all elements of $\sigma$, and all other hyperparameter are as shown in table \ref{alg:sNES}). The acronym (f.r.) indicates that this is the full range we tested, so the complete acceptable range may be larger. 

\begin{center}
\begin{tabular}{ |c|c|c| } 
 \hline
  & Generalized Cluster \\
  \hline
  $k$ & 10-40 (f.r.)  \\
  \hline
  $\eta_\sigma$ & 1e-1 - 1e-3 (f.r.)  \\
  \hline
  $\eta_\theta$ & 1 - 2e-2 \\
  \hline
  $\eta_r$ & 1e-2 - 1e-6 (f.r.) \\
  \hline
  $\sigma_{0}$ & 5e-1 - 1e-2 \\
 \hline
\end{tabular}
\end{center}

\subsection{Regularization} \label{sec:reg}

In this work we use two different types of regularization. The first is analogous to $L_2$ regularization in classical machine learning. Because this regularization is a common and simple yet effective choice for a wide variety of machine learning, it is appropriate in the absence of a more specific objective. Because each step of our ES optimization is intended to be similar to a gradient step in gradient descent, we will use the derivative of the $L_2$ norm (removing any constant terms, as these will be absorbed into the hyperparameter $\eta_r$ shown in table \ref{alg:sNES}):
\begin{equation*}
    \text{reg}_1(\sigma) = c \nabla_\sigma \|\sigma\|_2^2 = \sigma
\end{equation*}

The second form of regularization is intended to specifically maximize certified area (see Equation \ref{eq:cert_volume}). Similar to above, we will simply take the derivative of the certified area. However, because the derivative of this area directly is more complicated, we use the natural log of the area:
\begin{equation*}
    \text{reg}_2(\sigma) = c \nabla_\sigma \ln \left ( \frac{2\pi^{D/2}}{D\Gamma(D/2)}\prod_{i=1}^D s_e \sigma_i \right ) = c \nabla_\sigma \sum_{i=1}^D \ln(\sigma_i) = \frac{1}{\sigma}
\end{equation*}

As to what type of regularization is best to use, it depends on the type of $\delta$ perturbations a practitioner expects to encounter. Despite the theoretical pros and cons of each approach, they seemed to perform comparably according to our metrics. We only saw significant differences between the two during relatively easy tasks we tried prior to the experiments presented in this paper where parameters could be perturbed extremely without affecting performance at all. In this case the $L_2$ regularization would tend to increase the perturbations to extreme levels, whereas the area regularization would produce more moderate results.

\subsection{Probability Estimation} \label{sec:prob_est}

When in practice using our method outlined in Section \ref{sec:method} and Theorem \ref{indep_vars_thm}, it is impossible to exactly evaluate $p_A$ and $p_B$. This is because doing so would require you to compute the expectation over a Gaussian on the PQC. As such, when practically using this method, one must estimate the values of $p_A$ and $p_B$. To do so in such a way that still guarantees the conditions of theorem \ref{indep_vars_thm} hold, we can use some form of concentration inequality or confidence interval method. These mathematical methods will give us values $\underline{p_A}, \overline{p_B}$ such that $p_A \geq \underline{p_A}$ and $p_B \leq \overline{p_B}$ with probability $1 - \delta'$. Note that theorem \ref{indep_vars_thm} still holds if we replace $p_A$ with a lower bound on the true value, and likewise if we replace $p_B$ with an upper bound on the true value. As such, the whole theorem can still be used if these probabilities are estimated via sampling. And the more sampling a practitioner does, the better their estimate will become, and as a result also their robust certificate. In practice one can use whichever concentration inequality or confidence interval method they prefer, but in most cases the Clopper-Pearson confidence interval is used \citep{cohen_certified_2019, tecot_robustness_2021}.

\section{Theorem \ref{indep_vars_thm} Proof} 

For convenience we restate theorem \ref{indep_vars_thm} here:

\begin{theorem} [Noise-resilient Condition for Smoothed PQC Classifier] 
    Let $C$ be a PQC classifier and $G_{\sigma}$ to be the corresponding smoothed PQC classifier. If $G_{\sigma}(\theta, x) = c_a$, then $G_{\sigma}(\theta+\delta, x) = c_a$ for any $\delta$ vectors that satisfy
    \begin{align}
        \| \delta \oslash \sigma \|_2 &< \frac{1}{2} \big (\Phi^{-1}(p_A) - \Phi^{-1}(p_B) \big )
    \end{align} 
     where $\oslash$ is the Hadamard (element-wise) division, $\|\cdot\|_2$ is the $L_2$ norm, $\Phi^{-1}$ is the inverse of the standard Gaussian CDF, and 
    \begin{align}
        p_A &= \PP \left ( \argmax_{i \in \gamma} [C(\theta+\epsilon, x)_i] = c_a \right ) \\
        p_B &= \max_{c\neq c_a} \PP \left ( \argmax_{i \in \gamma} [C(\theta+\epsilon, x)_i] = c \right )
    \end{align} 
    with $\epsilon \sim \mathcal{N}(0, \Sigma)$ and $\Sigma$ is a diagonal matrix with vector $\sigma^{2}$ as the diagonal.
\end{theorem}

To assist in our proof, we will first define and prove:

\begin{lemma} \label{lemma_4}

Let $X \sim \mathcal{N}(x, \Sigma)$ and $Y \sim \mathcal{N}(x + \delta, \Sigma)$. Let $h : \RR^d \rightarrow \{0, 1\}$ be any deterministic or random function. Then:

\begin{enumerate}
    \item If $S = \{ z \in \RR^d : \lambda^T z \leq \beta \}$ for some $\beta$ and $\PP(h(X) = 1) \geq P(X \in S)$, \\ then $\PP(h(Y) = 1) \geq P(Y \in S)$
    \item If $S = \{ z \in \RR^d : \lambda^T z \geq \beta \}$ for some $\beta$ and $\PP(h(X) = 1) \leq P(X \in S)$, \\ then $\PP(h(Y) = 1) \leq P(Y \in S)$
\end{enumerate}

Where $\lambda = \delta \oslash \sigma^{\circ 2}$. ($\oslash$ is hadamard division, $\cdot^{\circ 2}$ is element-wise square.)

\end{lemma}

\emph{Proof}. 
This lemma is the special case of Lemma 3 in \cite{cohen_certified_2019} when $X$ and $Y$ are Gaussians with means $x$ and $x + \delta$. By Lemma 3 in \cite{cohen_certified_2019} it suffices to simply show that for any $\beta$, there is some $t > 0$ for which:
\begin{align}
    \{z : \delta^T z \leq \beta \} = \{ z : \frac{\mu_Y(z)}{\mu_X(z)} \leq t\} \text{ \quad and \quad} \{z : \delta^T z \geq \beta \} = \{ z : \frac{\mu_Y(z)}{\mu_X(z)} \geq t\}
\end{align}
The likelihood ratio for this choice of $X$ and $Y$ is:
\begin{align}
    \frac{\mu_Y(z)}{\mu_X(z)} &= \frac{\exp(\sum^{d}_{i=1} \frac{-1}{2\sigma_i^2} (z_i - (x_i + \delta_i))^2)}{\exp(\sum^{d}_{i=1} \frac{-1}{2\sigma_i^2} (z_i - x_i)^2)} \\
    &= \exp(\sum^{d}_{i=1} \frac{1}{2\sigma_i^2} (2z_i \delta_i - \delta_i^2 - 2 x_i \delta_i)) \\
    &= \exp(\lambda^\top z + b)
\end{align}
Where $\lambda = \delta \oslash \sigma^{\circ 2}$ and $b = \frac{-1}{2}\|\delta \odot \lambda\|_1 - \|x_i \odot \lambda\|_1$. ($\oslash$ is hadamard division, $\cdot^{\circ 2}$ is element-wise square.)

Therefore, given any $\beta$ we may take $t = \exp(\lambda^\top z + b)$, noticing that:
\begin{align}
    \lambda^\top z \leq \beta \Longleftrightarrow \exp(\lambda^\top z + b) \leq t \\
    \lambda^\top z \geq \beta \Longleftrightarrow \exp(\lambda^\top z + b) \geq t
\end{align}
\textbf{Finally, we can prove Theorem \ref{indep_vars_thm}.}

To show that $G_{\sigma}(\theta + \delta, x) = c_a$, it follows from the definition of $G_{\sigma}$ that we need to show that:
\begin{align}
    \PP(\argmax_{i \in \gamma} [C(\theta+\delta+\epsilon, x)_i] = c_a) \geq \max_{c_b\neq c_a} \PP(\argmax_{i \in \gamma} [C(\theta+\delta+\epsilon, x)_i] = c_b)
\end{align}
To show this, fix one class $c_b$ w.l.o.g. And for convenience we'll define the random variables:
\begin{gather}
    T := \theta + \epsilon = \mathcal{N}(\theta, \Sigma) \\
    Z := \theta + \delta + \epsilon = \mathcal{N}(\theta + \delta, \Sigma)
\end{gather}
We will choose any $\underline{p_A}$ and $\overline{p_B}$ such that:
\begin{align}
    \PP(\argmax_{i \in \gamma} [C(T, x)_i] = c_a) \geq \underline{p_A} \\
    \PP(\argmax_{i \in \gamma} [C(T, x)_i] = c_b) \leq \overline{p_B}
\end{align}
Given this, our goal is to show that 
\begin{align}
    \PP(\argmax_{i \in \gamma} [C(Z, x)_i] = c_a) > \PP(\argmax_{i \in \gamma} [C(Z, x)_i] = c_b)
\end{align}
To prove this, we define the half-spaces:
\begin{align}
    A :=&  \{ z : \lambda^\top (z - \theta) \leq \|\sigma \odot \lambda\|_2 \Phi^{-1}(\underline{p_A}) \} \\
    B :=& \{ z : \lambda^\top (z - \theta) \geq \|\sigma \odot \lambda\|_2 \Phi^{-1}(1 - \overline{p_B}) \} \\
\end{align}
Where $\lambda = \delta \oslash \sigma^{2}$, such that we can apply Lemma \ref{lemma_4} later ($\odot$ and $\oslash$ are the element-wise product and division respectively).

It can be shown that $\PP(T \in A) = \underline{p_A}$ (see section \ref{subsec:deffered_algebra}). Therefore, we know that $\PP(\argmax_{i \in \gamma} [C(T, x)_i] = c_a) \geq \PP(T \in A)$. Hence we may apply Lemma \ref{lemma_4} with $h := \textbf{1}[\argmax_{i \in \gamma} [C(z, x)_i] = c_a]$ to conclude:
\begin{align}
    \PP(\argmax_{i \in \gamma} C(Z, x)_i = c_a) \geq \PP(Z \in A)
\end{align}
Similarly, algebra shows that $\PP(T \in B) = \overline{p_B}$ (see section \ref{subsec:deffered_algebra}). Therefore, we know that $\PP(\argmax_{i \in \gamma} [C(T, x)_i] = c_b) \leq \PP(T \in B)$. Hence we may apply Lemma \ref{lemma_4} with $h := \textbf{1}[\argmax_{i \in \gamma} [C(z, x)_i] = c_b]$ to conclude:
\begin{align}
    \PP(\argmax_{i \in \gamma} [C(Z, x)_i] = c_b) \leq \PP(Z \in B)
\end{align}
Now all that is required is to show that $\PP(Z \in A) > \PP(Z \in B)$, as this implies:
\begin{align}
    \PP(\argmax_{i \in \gamma} [C(Z, x)_i] = c_a) \geq \PP(Z \in A) > \PP(Z \in B) \geq \PP(\argmax_{i \in \gamma} [C(Z, x)_i] = c_b)
\end{align}
We can compute that (shown in section \ref{subsec:deffered_algebra}):
\begin{align}
    \PP (Z \in A) &= \Phi (\Phi^{-1}(\underline{p_A}) - \frac{\langle \lambda, \delta \rangle}{\|\sigma \odot \lambda\|_2 }) \\
    \PP (Z \in B) &= \Phi (\Phi^{-1}(\overline{p_B}) + \frac{\langle \lambda, \delta \rangle}{\|\sigma \odot \lambda\|_2 })
\end{align}
Where $\langle \cdot, \cdot \rangle$ is the inner product of two vectors.

Using algebra we can determine when $\PP(Z \in A) > \PP(Z \in B)$, and $G(\theta+\delta,x) = c_a$ for all $\delta$ vectors that satisfy this inequality:
\begin{align}
    \PP(Z \in B) &< \PP(Z \in A) \\
    \Phi (\Phi^{-1}(\overline{p_B}) + \frac{\langle \lambda, \delta \rangle}{\|\sigma \odot \lambda\| }) &< \Phi (\Phi^{-1}(\underline{p_A}) - \frac{\langle \lambda, \delta \rangle}{\|\sigma \odot \lambda\| }) \\
    2\frac{\langle \lambda, \delta \rangle}{\|\sigma \odot \lambda\| } &< \Phi^{-1}(\underline{p_A}) - \Phi^{-1}(\overline{p_B}) \\
    \langle \lambda, \delta \rangle &< \frac{\|\sigma \odot \lambda\|}{2} (\Phi^{-1}(\underline{p_A}) - \Phi^{-1}(\overline{p_B})) \\
    \langle \delta, \delta \oslash \sigma^{\circ 2} \rangle &< \frac{\|\delta \oslash \sigma\|}{2} (\Phi^{-1}(\underline{p_A}) - \Phi^{-1}(\overline{p_B})) \\
    \| \delta \oslash \sigma \|^2 &< \frac{\|\delta \oslash \sigma\|}{2} (\Phi^{-1}(\underline{p_A}) - \Phi^{-1}(\overline{p_B})) \\
    \| \delta \oslash \sigma \| &< \frac{1}{2} (\Phi^{-1}(\underline{p_A}) - \Phi^{-1}(\overline{p_B})) \\
    \| \delta \oslash \sigma \| &< \frac{1}{2} (\Phi^{-1}(p_A) - \Phi^{-1}(p_B)) 
\end{align}

Note that in the last line we change $\underline{p_A} \rightarrow p_A$ and $\overline{p_B} \rightarrow p_B$ to illustrate the most favorable inequality possible. (But in practice one would have to do statistical sampling to estimate a $\underline{p_A}$ and $\overline{p_B}$ with high probability.)

\subsection{Deferred Algebra} \label{subsec:deffered_algebra}

\textbf{Claim.} $\PP(T \in A) = \underline{p_A}$

\textit{Proof.} Recall that $\sigma$ is a vector of standard deviations for each element, and $\Sigma$ is a diagonal matrix with $\sigma^{\circ 2}$ as the diagonal.
Additionally, note that $T \sim \mathcal{N}(\theta, \Sigma)$ and $A :=  \{ z : \lambda^\top (z - \theta) \leq \|\sigma \odot \lambda\| \Phi^{-1}(\underline{p_A}) \}$
\begin{align}
    \PP (T \in A) &= \PP (\lambda^\top (T - \theta) \leq \|\sigma \odot \lambda\| \Phi^{-1}(\underline{p_A}) ) \\ 
    &= \PP (\lambda^\top \mathcal{N}(0, \Sigma) \leq \|\sigma \odot \lambda\| \Phi^{-1}(\underline{p_A}) ) \\ 
    &= \PP (\|\sigma \odot \lambda\| \mathcal{N}(0, 1) \leq \|\sigma \odot \lambda\| \Phi^{-1}(\underline{p_A}) ) \\ 
    &= \PP (\mathcal{N}(0, 1) \leq \Phi^{-1}(\underline{p_A}) ) \\ 
    &= \Phi (\Phi^{-1}(\underline{p_A})) \\
    &= \underline{p_A}
\end{align}

\textbf{Claim.} $\PP(T \in B) = \overline{p_B}$

\textit{Proof.} Recall that $\sigma$ is a vector of standard deviations for each element, and $\Sigma$ is a diagonal matrix with $\sigma^{\circ 2}$ as the diagonal.
Additionally, note that $T \sim \mathcal{N}(\theta, \Sigma)$ and $B :=  \{ z : \lambda^\top (z - \theta) \geq \|\sigma \odot \lambda\| \Phi^{-1}(1 - \overline{p_B}) \}$
\begin{align}
    \PP (T \in B) &= \PP (\lambda^\top (T - \theta) \geq \|\sigma \odot \lambda\| \Phi^{-1}(1 - \overline{p_B}) ) \\ 
    &= \PP (\lambda^\top \mathcal{N}(0, \Sigma) \geq \|\sigma \odot \lambda\| \Phi^{-1}(1 - \overline{p_B}) ) \\ 
    &= \PP (\|\sigma \odot \lambda\| \mathcal{N}(0, 1) \geq \|\sigma \odot \lambda\| \Phi^{-1}(1 - \overline{p_B}) ) \\ 
    &= \PP (\mathcal{N}(0, 1) \geq \Phi^{-1}(1 - \overline{p_B}) ) \\ 
    &= 1 - \Phi (\Phi^{-1}(1 - \overline{p_B})) \\
    &= \overline{p_B}
\end{align}

\textbf{Claim.} $\PP (Z \in A) = \Phi (\Phi^{-1}(\underline{p_A}) - \frac{\langle \lambda, \delta \rangle}{\|\sigma \odot \lambda\| })$

\textit{Proof.} Recall that $\sigma$ is a vector of standard deviations for each element, and $\Sigma$ is a diagonal matrix with $\sigma^{\circ 2}$ as the diagonal.
Additionally, note that $Z \sim \mathcal{N}(\theta + \delta, \Sigma)$ and $A :=  \{ z : \lambda^\top (z - \theta) \leq \|\sigma \odot \lambda\| \Phi^{-1}(\underline{p_A}) \}$
\begin{align}
    \PP (Z \in A) &= \PP (\lambda^\top (Z - \theta) \leq \|\sigma \odot \lambda\| \Phi^{-1}(\underline{p_A}) ) \\ 
    &= \PP (\lambda^\top \mathcal{N}(0, \Sigma) + \langle \lambda, \delta \rangle \leq \|\sigma \odot \lambda\| \Phi^{-1}(\underline{p_A}) ) \\ 
    &= \PP (\|\sigma \odot \lambda\| \mathcal{N}(0, 1) \leq \|\sigma \odot \lambda\| \Phi^{-1}(\underline{p_A}) - \langle \lambda, \delta \rangle) \\ 
    &= \PP (\mathcal{N}(0, 1) \leq \Phi^{-1}(\underline{p_A}) - \frac{\langle \lambda, \delta \rangle}{\|\sigma \odot \lambda\| }) \\ 
    &= \Phi (\Phi^{-1}(\underline{p_A}) - \frac{\langle \lambda, \delta \rangle}{\|\sigma \odot \lambda\| })
\end{align}

\textbf{Claim.} $\PP (Z \in B) = \Phi (\Phi^{-1}(\overline{p_B}) + \frac{\langle \lambda, \delta \rangle}{\|\sigma \odot \lambda\| })$

\textit{Proof.} Recall that $\sigma$ is a vector of standard deviations for each element, and $\Sigma$ is a diagonal matrix with $\sigma^{\circ 2}$ as the diagonal.
Additionally, note that $Z \sim \mathcal{N}(\theta + \delta, \Sigma)$ and $B :=  \{ z : \lambda^\top (z - \theta) \geq \|\sigma \odot \lambda\| \Phi^{-1}(1 - \overline{p_B}) \}$
\begin{align}
    \PP (Z \in B) &= \PP (\lambda^\top (Z - \theta) \geq \|\sigma \odot \lambda\| \Phi^{-1}(1 - \overline{p_B}) ) \\ 
    &= \PP (\lambda^\top \mathcal{N}(0, \Sigma) + \langle \lambda, \delta \rangle \geq \|\sigma \odot \lambda\| \Phi^{-1}(1 - \overline{p_B}) ) \\ 
    &= \PP (\|\sigma \odot \lambda\| \mathcal{N}(0, 1) \geq \|\sigma \odot \lambda\| \Phi^{-1}(1 - \overline{p_B}) - \langle \lambda, \delta \rangle) \\
    &= \PP (\mathcal{N}(0, 1) \geq \Phi^{-1}(1 - \overline{p_B}) - \frac{\langle \lambda, \delta \rangle}{\|\sigma \odot \lambda\| }) \\ 
    &= \PP (\mathcal{N}(0, 1) \leq \Phi^{-1}(\overline{p_B}) + \frac{\langle \lambda, \delta \rangle}{\|\sigma \odot \lambda\| }) \\ 
    &= \Phi (\Phi^{-1}(\overline{p_B}) + \frac{\langle \lambda, \delta \rangle}{\|\sigma \odot \lambda\| })
\end{align}

\end{document}